\documentclass[epj]{svjour}

\usepackage{graphicx}
\begin{document}

\title{Continuous transfer and laser guiding between two cold atom traps}
\author{E.~Dimova\inst{1}\thanks{\email{Emiliya.Dimova@lac.u-psud.fr}}, O.~Morizot\inst{2}\thanks{\email{morizot@lpl.univ-paris13.fr}}, G.~Stern\inst{1}, C.L.~Garrido Alzar\inst{2}, A.~Fioretti\inst{1}, V.~Lorent\inst{2}, D.~Comparat\inst{1},
H.~Perrin\inst{2} and P.~Pillet\inst{1}} 
\institute{{Laboratoire Aim{\'e} Cotton, CNRS, B{\^a}timent 505, Universit\'e Paris Sud, F-91405 Orsay, FRANCE} \and {Laboratoire de physique des lasers, CNRS-Universit\'e Paris 13, 99 av. Jean-Baptiste Cl\'ement, F-93430 Villetaneuse, FRANCE}}

\titlerunning{Continuous transfer and laser guiding between two cold atom traps}
\authorrunning{E.~Dimova et al.}

\abstract{%
We have demonstrated and modeled a simple and efficient method to transfer atoms from a first Magneto-Optical Trap (MOT) to a second one. Two independent setups, with cesium and rubidium atoms respectively, have shown that a high power and slightly diverging laser beam  optimizes the transfer between the two traps when its frequency is red-detuned from the atomic transition. This pushing laser extracts a continuous beam of slow and cold atoms out of the first MOT and also provides a guiding to the second one through the dipolar force. In order to optimize the transfer efficiency, the  dependence of the atomic flux on the pushing laser parameters (power, detuning, divergence and waist) is investigated. The atomic flux is found to be proportional to the first MOT loading rate. Experimentally, the transfer efficiency reaches $70\,\%$, corresponding to a transfer rate up to $2.7\times10^8$\,atoms/s with a final velocity of 5.5~m/s. We present a simple analysis of the atomic motion inside the pushing--guiding laser, in good agreement with the experimental data. 
}

\date{\today}

\maketitle

\section{Introduction}

The realization of degenerate quantum gases requires the production of an initial dense and cold trapped atomic sample. The lifetime of the trapped atoms must be long enough to allow for appropriate evaporative cooling ramps, lasting up to several tens of seconds. A standard vapour Magneto-Optical Trap (MOT) setup cannot always satisfy this last condition because of the relatively high background pressure of the atomic vapour in the cell. The use of a dispenser~\cite{2001PhRvA..64b3402R} or of a desorption source~\cite{2001PhRvA..63b3404A,2003PhRvA..67e3401A} to load the MOT does not usually provide a trap lifetime longer than a few seconds. To obtain the required lifetime, the MOT has to be placed in an ultra-high vacuum chamber and loaded from a cold atom source, in general a slow and cold atomic beam. One of the demonstrated and widely used methods to create a cold atomic beam is the Zeeman slower technique. However, this solution requires an important technical development of different experimental techniques than the one implied in a MOT setup. In this paper, we will then concentrate on the transfer of atoms from a first cold source to a trap situated in a second high vacuum chamber.

There are several ways to transfer atoms from a cold atomic source to the high vacuum chamber. Mechanical devices~\cite{JILA} or magnetic guides~\cite{2001PhRvA...63:031401} are used to implement an efficient transfer of atoms initially in a MOT directly to either magnetic, electrostatic or atom chip traps. Other techniques, based on quasi-resonant light forces, allow a faster transfer to a recapture MOT.
Beam velocities low enough to allow the capture in a MOT in an ultra-high vacuum chamber can be obtained by the pyramidal MOT \cite{1998OptCo.157..303A,2003OSAJB..20.1161K}, the conical mirror funnel~\cite{2001PhRvA..64a3402K} or the two-dimensional MOT~\cite{1998PhRvA..58.3891D,2002PhRvA..66b3410S,2003OptCo.226..259C,2004PRL...93..093003}. Even simpler devices exist such as  the low velocity intense atomic source (LVIS)~\cite{1996PhRvL..77.3331L,2002EPJD...20..107C,2002OptCo.212..307T}.
Very high flux, up to $3 \times 10^{12}$~atoms/s, have been reported with a transversely cooled candlestick Zeeman slower type of setup~\cite{2004physics...7040S}. However, the counterpart of this large flux is a higher atomic velocity, $116$~m/s in this last experiment, which is by far too high to load a second MOT. A pulsed multiple loading, starting from a three-dimensional MOT, has been performed in Ref.~\cite{1996OptLett.21..290}. The atoms are pushed by a near resonant laser beam  resulting in a high number of atoms $1.5 \times 10^{10}$ in second MOT, loading rate $2 \times 10^8$ atoms/s and allow lower  velocity 16~m/s. However, the transfer is based on using an hexapole magnetic field, produced by a current above 60~A, which complicates the experiment. Simpler devices, without magnetic guiding, have achieved similar result by using continuous transfer~\cite{Wohlleben2001,Cacciapuoti2001,ChinesePhys}. In these experiments, a thin extraction column is created in the centre of a MOT and, due to a radiation pressure imbalance, a continuous beam of cold atoms is produced. It is possible to couple these simple devices with a distinct dipolar atomic guide~\cite{2000PhRvA..61c3411M}. We propose here to use the same laser beam for pushing and guiding the atoms, resulting in an even simpler setup.

This paper reports on a double MOT setup combining the ability of a pushing laser to extract the atoms  from a first trap (MOT1) and to guide them to a second trap (MOT2). The idea is to merge the leaking MOT technique ~\cite{1996PhRvL..77.3331L,Wohlleben2001,Cacciapuoti2001} with the red-detuned far off--resonance optical dipole guide technique~\cite{1999OptCo.166..199P,1999EL.....45..450S,2002NJPh....4...69W}. Two experiments have been simultaneously performed in two different laboratories, with different atoms: $^{133}$Cs at Laboratoire Aim{\'e} Cotton and $^{87}$Rb at Laboratoire de phy\-si\-que des lasers. Our setups are as simple as the one used in the leaking MOT techniques, but provide a higher flux and a lower atomic beam velocity. We can achieve a transfer efficiency up to $70\,\%$ with a mean atomic velocity 4 to 12 m/s depending on the pushing beam parameters. Our setups are very robust against mis-alignments of the pushing and guiding laser beam, and small variations of its detuning or power. The only requirement is a sufficiently high laser power (tens of mW) to produce a significant dipolar force to guide the atoms during their flight.

This paper is organized as follows: in section 2 we give details on the experimental realization of the beam and discuss the role of MOT1 parameters. In section 3 we describe theoretically the pushing and guiding processes during the atom transfer. Section 4 discusses the experimental parameter dependences of the setup as compared with the theory. Finally we present a comparison with other available techniques.

\section{Experimental realization}

\subsection{Experimental setup}

The vacuum system is similar in both experiments, except for a slight difference in the design of the differential vacuum tubes and the MOT2 cells.

\begin{figure}[ht]
\begin{center}
\resizebox{0.3\textwidth}{!}{
\includegraphics*[width=\linewidth]{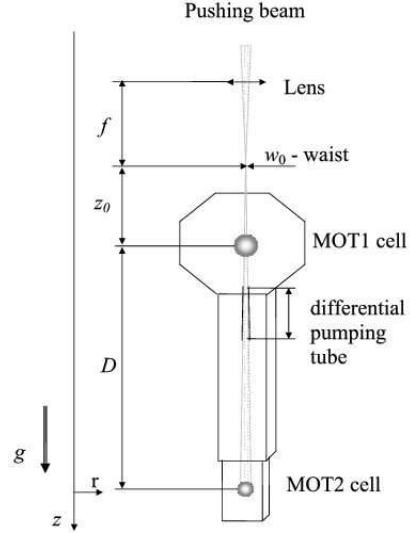}}
\caption{Scheme of the experimental setups. The parameters used in the discussion ($f$, $D$, $z_0$, $w_0$) are labeled on the picture. The vertical $z$ axis is oriented downwards.}
\label{fig:setup}
\end{center}
\end{figure}

For the cesium (resp. rubidium) experiment the setup consists of two cells vertically separated (see Figure~\ref{fig:setup}). The distance between the two traps is $D=57$~cm (resp. $D=72$~cm). A reservoir connected to the upper source cell supplies the atomic vapour. The recapture chamber is a glass cell with $1\times1\times10$~cm$^3$ (resp. $1.25 \times 7.5\times 12$~cm$^3$) volume. A differential pumping tube located $3$~cm (resp. 10~cm) below MOT1 provides a  vacuum within the $10^{-11}$~mbar range in the bottom MOT2 cell while in the MOT1 cell it is in the $ 10^{-8}-10^{-9}$\,mbar range. For the cesium experiment, the tube is $18$~cm long and has a conical shape ($3$~mm diameter at its top and $6$~mm at its bottom part) whereas it is cylindrical, 12~cm long and 6~mm diameter in the rubidium experiment.

In both cases, MOT1 runs in a standard magneto-optical trap configuration with a magnetic field gradient around $15$\,G/cm along the horizontal axis of the MOT1 coils. All the laser beams have a $2.5$~cm diameter (clipped by the  mounts of the quarter-wave plates) and are provided by laser diodes. In the rubidium experiment, the laser is divided into 3 retroreflected beams carrying 10~mW laser power. They are 10~MHz red-detuned from the $^{87}$Rb $5s(F=2)\rightarrow 5p_{3/2}(F'=3)$ transition. In the cesium experiment, two $5$\,mW radial beams are retroreflected and make an angle $\pm 45^{\circ }$ with the vertical axis. Each of the two (non reflected) axial beams carries $10$\,mW laser power. They are $15\,$MHz red-detuned from the Cs $6s(F=4)\rightarrow 6p_{3/2}(F'=5)$ transition. The 5~mW repumping light, with a frequency on resonance respectively with the Cs transition $6s(F=3)\rightarrow 6p_{3/2}(F'=4)$ and the $^{87}$Rb transition $5s (F=1) \rightarrow 5p_{3/2}\, (F'=2)$, is mixed with all the trapping beams. In MOT2, the trapping beams are limited to about $2R=8$~mm in diameter in both experiments due to the cell dimensions and in order to reduce the scattered light.

In addition to these trapping lasers, the linearly polarized pushing--guiding beam, red-detuned from the MOT ($F \longrightarrow F+1$) transition ($F=4$ for Cs, $F=2$ for $^{87}$Rb) with maximum power of $P_0 = 63$~mW for Cs (resp. $P_0 = 21$~mW for Rb), is aligned vertically into the trap. The parameters used in both experiments are summarized in Table~\ref{tab:param}. In contrast with the similar setups reported in \cite{Wohlleben2001,Cacciapuoti2001}, the pushing lasers are not frequency-locked in our experiments. The detuning is chosen to optimize the transfer efficiency and is found to be such that the number of atoms in MOT1 is roughly reduced by a factor ten when the ``pushing--guiding beam'' is present. The beam is focused at position $z_0=-34$\,cm (resp. $z_0=-13$~cm) before MOT1 by a lens $f=2$\,m (resp. $f=1$~m). It is not perfectly Gaussian, however the waist at position $z$ is still given by $w(z)=w_{0}\sqrt{1+(z-z_{0})^{2}/z^{2}_{R}}$, where $w_{0}=200~\mu$m (resp. $300~\mu$m) is the measured minimum waist and $z_R= 110$~mm is the estimated Rayleigh length (resp. $z_R=260~$~mm, measured value for Rb). It diverges to  a $1/e^2$-radius of $w_1=0.65$\,mm (resp. 0.33~mm) in MOT1 and about $1.7$\,mm (resp. 1.0~mm) in MOT2. The larger size of the beam at the position of MOT2 limits the perturbation of the trapping and cooling mechanisms.

\begin{table}
\caption{Pushing beam parameters used in cesium and rubidium experiments (see text and Figure~\ref{fig:setup}). All distances are given in mm, the laser power $P$ is in mW. $w_0$, $w_1$ and $w_2$ are the pushing beam radius at $1/e^2$ at focus, MOT1 and MOT2 positions, respectively.}
\label{tab:param}       
\begin{tabular}{lllllllll}
\hline\noalign{\smallskip}
exp. & $D$ & $f$ & $|z_0|$ & $w_0$ & $w_1$ & $w_2$ & $z_R$ & $P$  \\
\noalign{\smallskip}\hline\noalign{\smallskip}
Cs & 570 & 2000 & 340 & 0.2 & 0.65 & 1.7 & 110 & $< 63$\\
Rb & 720 & 1000 & 130 & 0.3 & 0.33 & 1.0 & 260 & $< 21$\\
\noalign{\smallskip}\hline
\end{tabular}
\end{table}

\subsection{Flux from MOT1}
Experimentally, the main features of the atomic beam are deduced from the loading characteristics of MOT1 and MOT2, where the number of atoms is determined using a calibrated photodiode monitoring the scattered MOT light.
The main goal is to have the highest possible recapture rate of atoms in the MOT2 region. This ingoing flux is obviously related to the characteristics of MOT1.

The extraction process can be summarized as follows \cite{Wohlleben2001,Cacciapuoti2001}. In MOT1 hot atoms are first decelerated by the MOT radiation pressure, then slowly moved to the centre of the trap where they are extracted by the pushing laser. In addition to its pushing effect, the laser beam shifts the atomic levels by a few natural linewidths so that a transverse  cooling of the atomic beam takes place during extraction, limiting the initial atomic temperature to about $25~\mu$K for Cs ($40~\mu$K for Rb). Moreover, the trapping forces are reduced and the pushing beam becomes dominant. Hence, atoms are extracted from the trap and accelerated in the direction of MOT2. After the transfer to the second chamber, the atoms are finally recaptured in MOT2 by radiation pressure.

In a first set of experiments, we study the flux of atoms extracted from the upper chamber. This outgoing flux depends on the number of captured atoms in MOT1, which is related to the background pressure of the alkali vapour. As there is no direct access to the background pressure value, we have measured the loading time of MOT1, which, in a large regime of operating parameters, is inversely  proportional to the atomic pressure in the source cell. The number of atoms in a MOT in the stationary regime is~\cite{Steane92}
	\begin{equation}
	N=\frac{L}{\gamma+\gamma_{p}+\beta n} ,
	\end{equation}
where L represents the loading rate of the MOT, $\gamma $ is the loss rate due to background collisions, $\gamma_p$  gives the loss rate induced by the pushing laser, $\beta $ is the rate of the cold two-body collisions between the trapped atoms, and $n$ is the average atomic density in the MOT. The density in MOT1 is limited to about $10^{10}$~atoms/cm$^3$, so that the term $\beta n$ is negligible in both setups. 
 
\begin{figure}[t]
\includegraphics*[width=\linewidth]{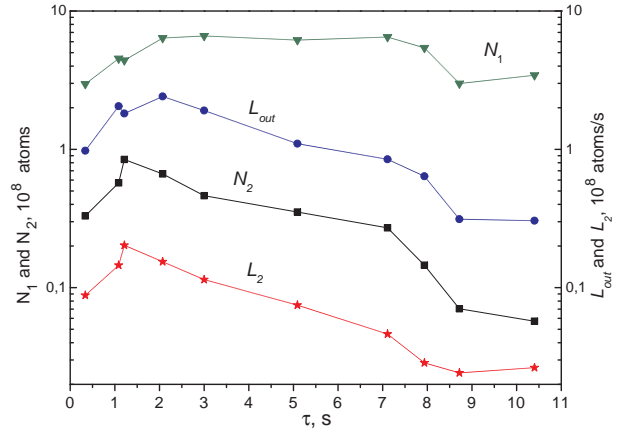}
\caption{Dependence of the MOT2 parameters on the MOT1 loading time $\tau$ in the Cs experiment. $N_{1}$ is the number of atoms in MOT1 without pushing beam, $L_{\rm out}$ is the atomic flux, $L_2$ the loading rate of MOT2 and $N_{2}$ the number of recaptured atoms in MOT2.}
\label{fig:fluo2}
\end{figure}

Loading rates in both MOTs are given by the measured initial slope of the number of trapped atoms in the MOT versus time. In MOT2, this measure is performed after suddenly switching on the pushing laser and waiting for the arrival of the first atoms. In MOT1, the loss rate $\gamma$ is inferred by measuring the 1/$e$-loading time $\tau$ ($\gamma=1/\tau$ in a wide range of vapour pressure~\cite{Steane92}) or by dividing the loading rate $L_1$ by the number of atoms $N_1$ measured when the pushing beam is off. When the pushing laser is switched on, the loss processes in MOT1 increase drastically. If $N^p_1$ is the number of atoms in MOT1 in the presence of the pushing beam, then $L_{\rm out}=\gamma_p N^p_1$ is the flux of atoms leaking out of MOT1 through the optical guide. We deduce it from parameters we already measured via the formula:
\begin{equation}
	L_{\rm out}=L_1-\gamma N^p_1.
\end{equation}

To get the data plotted on Figure~\ref{fig:fluo2}, $\tau$ is tuned by varying the background pressure. Whatever the background pressure, the number $N_{1}$ of atoms is approximately constant. At  high background pressure ({\em i.e.} low values of $\tau$), the outgoing atomic flux $L_{\rm out}$ increases with the loading time $\tau$ because the number of atoms without pushing beam $N_{1}$ slightly does. Then at relatively low  pressure $L_{\rm out}$ decreases, following the behaviour of the MOT1 loading rate $L_1$ (inversely proportional to $\tau$). The loading rate of MOT2 $L_2$  and the number of atoms $N_{2}$ in MOT2 are presented as a function of $\tau$ on Figure~\ref{fig:fluo2}. Their dependence with the MOT1 loading time is similar to that of the atomic flux $L_{\rm out}$. The overall efficiency of the transfer process is defined by the incoming flux in MOT2 divided by the outgoing flux from MOT1, that is $L_2/L_{\rm out}$.

We conclude that for higher MOT2 loading rate we need a relatively high background pressure in MOT1 and a large laser power in the trapping beams (to have higher $N_{1}$ value). For our experimental conditions the optimum is at a MOT1 loading time of about 1-2~s. The data presented here were not taken with optimized pushing--guiding beam parameters, the efficiency being here limited to typically 10\%. Once these parameters are well set we are able to achieve maximum transfer efficiency of about $70\%$ for Cs (resp. $50\%$ for $^{87}$Rb), without affecting the overall dependence of the different quantities on the MOT1 loading time.

\section{Pushing and guiding processes}

After leaving the MOT1 region, the atomic beam is no longer affected by the MOT1 lasers and is guided due to the attractive dipolar force created by the red-detuned pushing--guiding beam. In this section, we describe the guiding process using an analytical model similar to that given in references \cite{1999OptCo.166..199P,2002NJPh....4...69W}. The total force applied on the atoms is the sum of a radiation pressure ``pushing force'' $ \vec F_{\rm push} $ and of a dipolar ``guiding force'' $ \vec F= - \vec \nabla U$, where U is the guiding potential. The gravitational force plays a minor role in the loading process. 

A two-level model function of the laser parameters (power, detuning and waist) describes qualitatively the experimental dependence of the transfer efficiency between the two MOTs. A more detailed quantitative analysis of the processes is proposed in the rest of this section.

\subsection{Two-level model}
\label{subsec:2level}

In this first simple model we neglect gravity, the initial velocity and temperature of the atoms and beam divergence. We consider the atoms as a two-level system with a transition energy $\hbar \omega_0$, a natural linewidth $\Gamma$ ($\Gamma/2\pi = 5.2$~MHz for Cs, 5.9~MHz for Rb) and a saturation intensity $I_s = \frac{1}{6} \hbar c k^3 \frac{ \Gamma}{ 2 \pi} $ (1.1~mW/cm$^2$ for Cs, 1.6~mW/cm$^2$ for Rb). We use here $z$ as the vertical coordinate along the laser beam propagation with origin in the centre of MOT1 and $r$ for the radial cylindrical coordinate (see Figure~\ref{fig:setup}). For this two-level model, the waist $w$ of the pushing--guiding laser is assumed to be constant and equal to its experimental value at MOT1 position $z=0$. The laser beam has a power $P_0$, a wave vector $k = 2\pi/\lambda $, and an angular frequency $\omega$ detuned by $\delta = \omega - \omega_0$ with respect to the atomic transition.
 
The on-axis light shift is given by $U_{0} = \frac{\hbar \delta}{2}  \ln( 1 + s ) $ where $s = (I/I_s)/(1 + 4 \delta^2/\Gamma^2)$ is the saturation parameter and $I= 2 P_0/(\pi w^2)$ is the peak laser intensity. As the laser is far detuned, saturation is always very low and one can simply replace $\ln(1+s)$ by $s$ in this expression. In this limit, the guiding potential is 
\begin{equation}
	 U_{0} \, e^{-\frac{2r^{2}}{w^{2}}} \, .
\end{equation}
As the waist is considered constant, the guide does not affect the longitudinal motion. On the contrary, it is crucial for confining the transverse motion.

The atoms absorb and emit spontaneously photons at a rate 
\begin{equation}
	\Gamma' = \frac{\Gamma}{2} \frac{ s } {1+s} =  \frac{\Gamma}{2}\frac{I/I_s}{1+4\frac{\delta^2}{\Gamma^2}+I/I_s},
\end{equation}
 which gives a pushing force 
\begin{equation}
	 F_{\rm push}  = \Gamma' \hbar k  = \Gamma' M v_{\rm rec},
\end{equation}
where $v_{\rm rec}=\hbar k/M$ is the recoil velocity and $M$ the atomic mass. The velocity increases due to photon absorption, and the number of scattered photons to reach the position $z$ is approximately $v(z)/v_{\rm rec}=\sqrt{2\Gamma' z/v_{\rm rec}}$. The pushing process is also responsible for a heating due to random spontaneous emission in all directions. The mean horizontal kinetic plus potential energy per atom $2 k_B T$ in the 2D confining potential is increased by two third of the recoil energy $E_{\rm rec} = M v_{\rm rec}^2/2 = k_B T_{\rm rec}/2$ at each scattering event~\cite{Grimm99,2002NJPh....4...69W}. This gives rise to a horizontal kinetic temperature
\begin{equation}
T_h(z)= \frac{v(z)}{v_{\rm rec}} \frac{T_{\rm rec}}{6}.
\label{eqn:T_horizontal}
\end{equation}

To have an efficient pushing--guiding beam we require in this simple two--level approach that the atoms remain trapped in two dimensions inside the guide during the whole transfer. This condition is
\begin{equation}
2 k_{\rm B} T_h(z) < |U_0| \mbox{ for all }z.
\label{eqn:trapped}
\end{equation}
As the horizontal velocity spread increases with $z$, this is equivalent to $2 k_{\rm B} T_h(D) < |U_0|$. A second constraint is that the beam velocity at the MOT2 position ($v_{\rm beam}$) should be lower than the capture velocity ($v_{\rm capture}$) of the MOT 
\begin{equation}
v_{\rm beam} < v_{\rm capture}.
\label{eqn:velocity}
\end{equation}
The value of $v_{\rm capture}$ is on the order of the maximal velocity for an atom to be stopped on the MOT beam diameter distance $2R$, that is $v_{\rm capture}=\sqrt{\Gamma R v_{\rm rec}}$~\cite{Steane92}. As a result, we evaluate $v_{\rm capture}$ to be about 21~m/s for cesium and 30~m/s for rubidium.

\begin{figure}[t]
\begin{center}
\includegraphics[width=0.9\linewidth]{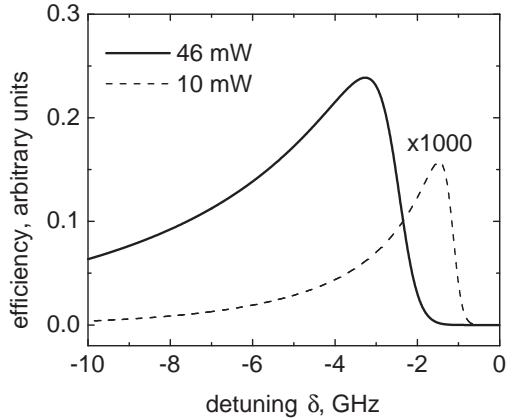}
\end{center}
\caption{Efficiency (see text) of the pushing--guiding processes versus laser detuning $\delta$ for a $650\,\mu$m waist laser beam with different laser power 10~mW (dashed line $\times 1000$) and 46~mW (solid line).
The other parameters are: initial temperature $T_0 = 25\,\mu$K,
 $I_{\rm sat}=1.1\,$mW/cm$^2$, $\Gamma = 2\pi \times 5.2\,$MHz, $v_{\rm rec} = 3.5\,$mm/s, $T_{\rm rec} =  0.2\,\mu$K and the two MOT cells are separated by $57\,$cm (the values used are those of the Cs experiment).}
\label{fig:pot_2_niv}
\end{figure}

The efficiency of the pushing--guiding process is determined by how deep the conditions~(\ref{eqn:trapped}) and~(\ref{eqn:velocity}) are verified. To describe qualitatively the guiding efficiency in relation with these conditions, we propose to describe each condition by a function $f$, equal to zero when the inequality is strongly violated and to 1 when it is fully verified, with a continuous transition between these two extremes. The guiding efficiency is then described by the product $f(\frac{2 k_{B}T_{h}(D)}{\left|U_{0}\right|}) \times f(\frac{v_{\rm beam}}{v_{\rm capture}})$ of the two conditional functions. The result is given for Cs in Figure~\ref{fig:pot_2_niv} as function of the laser detuning, with $v_{\rm capture}=21$~m/s and the function $f$ chosen arbitrarily to be $f(x)=\frac{1}{1+x^{10}}$.

A comparison of the two-level model with experimental results (see Figure~\ref{fig:detuning}, left) presents a good qualitative agreement, reproducing the presence of an optimal red detuning at given laser power. The maximum transfer efficiency increases with the power of the pushing beam while the position of the peak is shifted to larger absolute values of the detuning. This simple model is sufficient to derive the main conclusion: the transfer is more efficient with a far red-detuned and intense laser beam. However, the theory predicts a peak further from resonance than observed experimentally. Moreover, the sensitivity to the laser power is much more pronounced than observed in the experiment. This motivates a more detailed analysis of the processes operating during the travel of the atoms from MOT1 to MOT2. In particular, the effect of optical pumping to the lower hyperfine state has to be considered.

\subsection{Optical pumping}
\label{sec:OptPump}

The absorbed photons can lead to optical pumping between the two hyperfine levels of the ground state  which have different laser detuning with respect to the pushing laser. Indeed, very quickly after leaving the MOT1 region, the atoms are pumped essentially in the lower ground state $F=3$ for cesium (resp. $F=1$ for rubidium) as there is no repumping laser light superimposed with the pushing laser beam. This optical pumping is essential for a good transfer efficiency, as it greatly reduces the final velocity of the atomic beam (see sections \ref{subsec:TransferEff} and \ref{sec:BeamVelocity}). However, a small population in the other ground state is still present, typically 1 to 3 percent for a linearly polarized beam, as we shall see~\cite{note1}. As the radiation pressure is much larger for atoms in the upper ground state (about 100 times larger for detuning values discussed here), even this small fraction plays a role and both ground state populations have to be taken into account for the estimation of the pushing force. On the contrary, the dipolar force may be estimated by assuming that the atoms are only in the lower ground state, as this force is only about 10 times smaller than in the upper ground state, which is 100 times less populated.
 
An estimate of the populations in the ground states is obtained by assuming an equal detuning for the transitions from the upper hyperfine ground state to all the hyperfine excited states. We define  an ``effective'' detuning $\bar{\delta} \approx \delta + \Delta'_{\rm HFS}/2$, where $\Delta'_{\rm HFS}$ is the total width of the hyperfine structure in the excited state ($\Delta'_{\rm HFS}\simeq 600$~MHz for Cs and $\Delta'_{\rm HFS} \simeq 500$~MHz for $^{87}$Rb respectively). Using this mean detuning $\bar{\delta}$ we calculate the pumping rates between the two hyperfine ground states. This is fairly good for large detunings (above 1 GHz from the cycling transition). To illustrate our results we will choose the following typical values: $\delta/2 \pi =- 2$~GHz from the (F=4$\to$F'=5) transition of the Cs (\textit{i.e.} $\bar{\delta}/2 \pi = - 1.70$~GHz) and $\delta/2 \pi = - 1$~GHz (\textit{i.e.} $\bar{\delta}/2 \pi = - 750$~MHz) from the (F=2$\to$F'=3) transition of the $^{87}$Rb.
We also define
$\Delta_{\rm HFS}$ as the hyperfine structure interval in the ground state
($2\pi \times 9.2$~GHz for Cs, $2\pi \times 6.8$~GHz for $^{87}$Rb) (See~\cite{D.Steck}). 
 
The ratio of populations in the upper hyperfine  ground state $N_{F+1}$ and in the lower one $N_F$ may  then be estimated as:
\begin{equation}
\eta =  \frac{N_{F+1}}{N_{F}+ N_{F+1}} \approx \frac{N_{F+1}}{N_{F}} = \alpha \left( \frac{ \bar{\delta} }{ \bar{\delta} - \Delta_{\rm HFS} } \right)^2 
\end{equation}

with $\alpha = \displaystyle \frac{2F+3}{2F+1} = \left\{ 
\begin{array}{l} 9/7 \, \mbox{ for Cs } (F=3)\\
5/3 \, \mbox{ for $^{87}$Rb } (F=1)
\end{array} \right.$.

The factor $\alpha$ is simply the ratio between the number of substates in the $F+1$ and $F$ ground states, to which $N_{F+1}/N_F$ should be equal at a detuning large as compared to the hyperfine structure $\Delta_{\rm HFS}$; the term involving the detuning is related to the ratio of excitation rates from the two hyperfine ground states. The formula leads to $\eta = 3.2$~\% of the atoms in the Cs($6s, F=4$) state and  $\eta = 1.6$~\% in the Rb($5s, F=2$). This value is in excellent agreement with a full calculation taking into account all the different detunings with the hyperfine excited states.
 
\subsection{Pushing force}
\label{sec: Pushing force}
 
Another factor should be considered: the laser mode shape. Indeed, a relatively strong divergence is needed in order to both efficiently push and guide atoms in the MOT1 region and not affect the MOT2 operation. The guiding beam waist varies with position, according to
\begin{equation}
w(z)=w_{0}\sqrt{1+(z-z_{0})^{2}/z^{2}_{R}} \, .
\end{equation}
The depth $U_0$ is then modified along the atomic trajectory due to the change in the laser intensity and, taking into account the results of the previous section, the pushing force in the centre of the beam may be estimated as follows:
\begin{eqnarray}
	F_{\rm push}(z) &=&  \frac{\Gamma}{2} \hbar k \bar{s}(z) \left( (1 - \eta) + \eta \left( \frac{ \bar{\delta} - \Delta_{\rm HFS} }{ \bar{\delta} } \right)^2 \right) \nonumber \\
	& & \simeq  \frac{\Gamma}{2} \hbar k \bar{s}(z) (1 + \alpha),
\end{eqnarray}
where $\bar{s}(z)$ is the saturation parameter calculated for the lower ground state, at detuning $\bar{\delta} - \Delta_{\rm HFS}$. We take into account the linear polarization of the pushing beam by multiplying $\bar{s}$ by a factor 2/3 in all calculations. We neglect however the small change in $\bar{\delta}$ due to the light shift, which is even reduced when the waist $w(z)$ becomes larger.

The mean pushing force is reduced due to the Gauss\-ian transverse profile of the pushing beam and the finite size of the atomic cloud. This may be taken into account approximately by dividing the force by a factor 2~\cite{2002NJPh....4...69W}. Note that this underestimates the initial pushing force, when the atoms are still well guided (rms radius less than $w(z)$/2), and overestimates it when the cloud size becomes larger than half the waist. As the mean pushing force now depends only on $z$, it may be written as the derivative of a ``pushing potential'' $U_{\rm push}$:
\begin{equation}
U_{\rm push}(z) = \frac{\Gamma}{2} \hbar k s_0 z_R (1 + \alpha) \, \arctan \frac{z-z_0}{z_R} .
\end{equation}
where $s_0=\bar{s}(z_0)$. The velocity at each point is then easily calculated by energy conservation ($z$ axis is oriented downwards):

\begin{eqnarray}
\lefteqn{	v(z)= \Big[ v_0^2 + 2 g z + }	\label{eqn:vel} \\
	& & \left. \Gamma v_{\rm rec} s_0 z_R (1 + \alpha) \, \left( \arctan \frac{z-z_0}{z_R} + \arctan \frac{z_0}{z_R} \right) \right]^{1/2} ,
\nonumber
\end{eqnarray}
where $v_0$ is the input velocity in the guide. The effect of gravity is not dominant, but was taken into account by the $2 g z$ term. $v_0$ can be estimated as the output velocity of the MOT1 region. We have calculated it using formula~(\ref{eqn:vel}) assuming that the atoms in the MOT1 region  have a zero initial velocity and are in the upper hyperfine ground state due to the presence of the repumping light ($\eta=1$). For instance, using a travel distance $z$ roughly equals to the MOT1 region radius (10~mm) and a laser power of 21~mW, we find that atoms enter the guide with a velocity $v_0\approx9$ m/s for Rb; for the Cs parameters, we obtain in the same way $v_0\approx3.1$~m/s. From Equation~(\ref{eqn:vel}), we also infer the traveling time as $\Delta t = \int_0^D \frac{d z}{v(z)} $.

\begin{figure*}
\includegraphics*[width=0.45\linewidth]{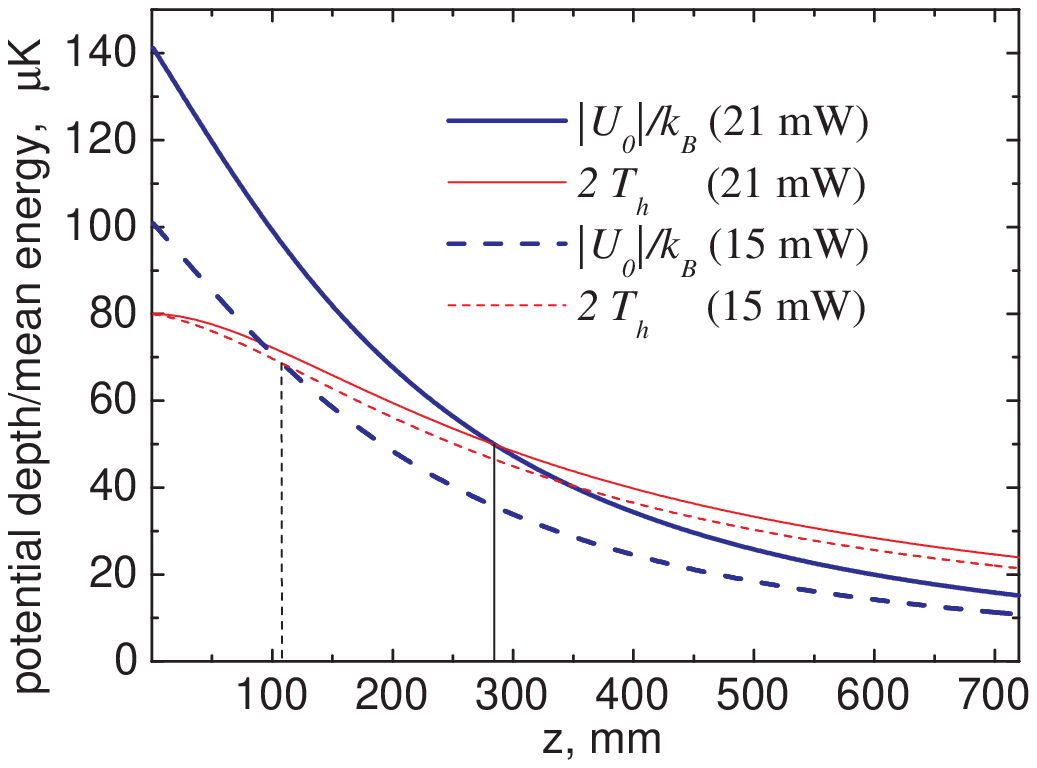}
\hspace{5mm}
\includegraphics*[width=0.48\linewidth]{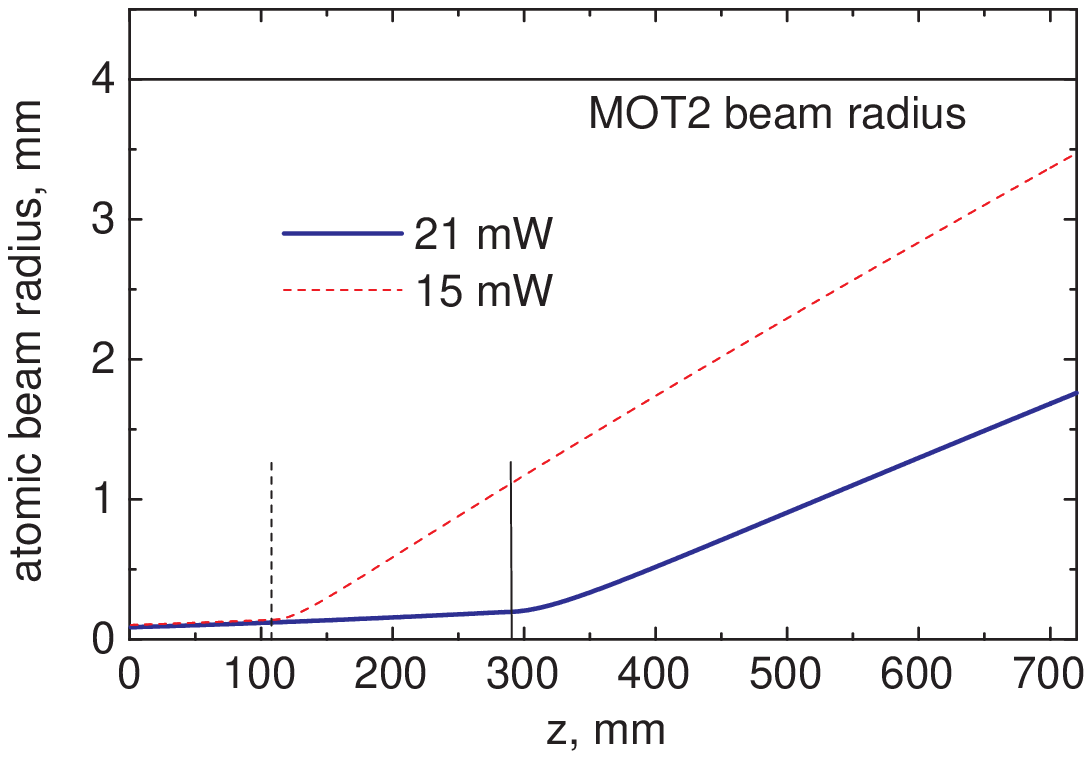}
\caption{Evolution for the $^{87}$Rb experiment of different parameters with the traveling distance $z$ between the two MOTs, at different powers: 15~mW (dashed lines) and 21 mW (solid lines); the initial temperature at the guide entrance was set to $T_0=40~\mu$K. The point $z_{\rm out}$ where the atoms leave the guide is marked by a vertical line. Left: mean horizontal energy $2 k_B T_h$ (thin lines, Equation~\ref{eqn:Th}) and trap depth $|U_0|$ (bold lines, Equation~\ref{eqn:depth}). Right: rms radius of the guided atomic beam. The radius of the MOT2 beams is marked by a horizontal line.}
\label{fig:sizeRb}
\end{figure*}

\subsection{Guiding condition}
\label{subsec: Guiding potential}

As previously discussed, the light shift of the lower ground state is dominant in our case. The atoms leaving MOT1 are thus guided by the on-axis light shift potential given by 
 
\begin{equation}
		U_{0} (z) =  \frac{\hbar (\bar{\delta} - \Delta_{\rm HFS})}{2} \, \bar{s}(z).
		\label{eqn:depth}
\end{equation}

Equation~(\ref{eqn:trapped}) is still the strongest constraint for the choice of the parameters and becomes more and more difficult to fulfil as $z$ increases, because $|U_{0}|$ is reduced due to the beam divergence. The horizontal kinetic temperature $T_h(z)$ is evolving due to two opposite effects: photon scattering~\cite{note2} is responsible for an increase of $T_h$ while adiabatic cooling tends to lower it as the waist increases. The adiabaticity condition $|d\omega_p/dt| \ll \omega_p^2$, where $\omega_p$ is the transverse oscillation frequency of the guide, is well fulfilled in both experiments except when the atoms move in the non harmonic part of the potential. This break-down of the adiabaticity occurs only when the atoms are close to leave the guide. This only marginally affects the guiding condition and will not be taken into account here. $\omega_p$ varies with the inverse squared waist, and one has $\omega_p(z)=\omega_p(0) w^2(0)/w^2(z)=\omega_p(0)\frac{z_0^2 + z_R^2}{(z-z_0)^2 + z_R^2}$. To obtain an expression for $T_h(z)$, valid while the atoms remain guided, we write the change in $T_h$ for a small change $\delta z$ in $z$.
As the phase space density is conserved during this adiabatic cooling, the cooling contribution is proportional to the inverse squared waist. Spontaneous scattering is responsible for a supplementary heating term, proportional to the number of photons scattered during $\delta t=\delta z/v$:
\begin{equation}
	T_h(z+\delta z) = T_h(z) \frac{w^2(z)}{w^2(z+\delta z)} + \frac{\Gamma}{2} \bar{s}(z) \frac{T_{\rm rec}}{6} \frac{\delta z}{v(z)} \, .
\end{equation}
The temperature increase is $T_{\rm rec}/6$ for each spontaneous scattering event. $\bar{s}(z)$ is proportional to $1/w(z)^2$, just like the oscillation frequency. Using the dependence in $w(z)$, we obtain the following differential equation for $T_h$:
\begin{equation}
	\frac{dT_h}{dz} = - T_h(z) \frac{2}{w(z)}\frac{dw}{d z} + \frac{T_{\rm rec}}{6}\frac{w^2(0)}{w^2(z)} \frac{\Gamma}{2} \bar{s}(0) \frac{1}{v(z)}
\end{equation}
Using the expression of $w(z)$, the solution of this equation reads:
\begin{equation}
T_h(z) = \frac{z_0^2 + z_R^2}{(z-z_0)^2 + z_R^2} \left[ T_0 + \frac{T_{\rm rec}}{6} \frac{\Gamma}{2} \bar{s}(0) \int_0^z\frac{dz'}{v(z')} \right]
\label{eqn:Th}
\end{equation}
where $T_0$ is the initial temperature at the guide entrance. The integral in the last term is the time necessary for an atom to travel to position $z$.
In the range of parameters explored in our experiments, the sum of these two terms decreases with $z$, but slower than the trap depth. As can be seen on Figure~\ref{fig:sizeRb} (left), the mean horizontal energy $2 k_B T_h$ becomes larger than the trap depth at some position $z_{\rm out}$ before reaching MOT2. However, as will be discussed below, this partial guiding is sufficient for limiting the size of the atom cloud to below the MOT2 beam diameter.

\subsection{Recaptured atoms}
\label{subsec:Recaptured}

For a good transfer efficiency, two main criteria have to be fulfilled. First, the atomic beam should stay roughly collimated on a distance long enough to pass through the differential tube, and then the transverse cloud radius at the end should be comparable to the capture radius of MOT2. This means that even if they leave the guide before reaching MOT2, the atoms can still be recaptured. Second, the final longitudinal velocity of the atomic beam must not exceed the capture velocity of MOT2. As the atomic beam velocity is in any case lower than the capture velocity of MOT2, the recapturing mechanism is mostly limited by the matching between the atomic beam size and the size of the capturing region of MOT2.

The capture size of MOT2 is limited by the radius $R=4$~mm of the collimated trapping laser beams. According to the former considerations about heating of the guided atoms (see~\ref{subsec: Guiding potential}), the mean horizontal energy of the cloud is lower than the guiding trap depth over a distance $z_{\rm out} = 38$~cm for a laser power of 63 mW and an initial temperature $T_0=25~\mu$K in the case of Cs (resp. $z_{\rm out} = 28.5$~cm with $T_0=40~\mu$K and a laser power of 21 mW in the case Rb) (see Figure~\ref{fig:sizeRb}). For simplicity we consider hereafter that all the atoms remain pushed and guided up to that point and then undergo a free ballistic expansion as they keep falling. Including this assumption in our model, we can evaluate the size of the atomic cloud $\Delta r_f$ as it reaches MOT2.  While the atoms remain trapped, the cloud size is of the order of $\omega^{-1}_p(z) \sqrt{k_B T_h/m}$. The guiding step ends when $k_B T_h$ reaches $|U_0(z)|/2$, such that the rms size at the guide output is $\Delta r_{\rm out} = \omega^{-1}_p(z)\sqrt{|U_0(z)|/2m} = w(z)/\sqrt{8}$, that is $\Delta r_{\rm out} = 470~\mu$m for Cs (resp. $\Delta r_{\rm out} = 200~\mu$m for Rb). We assume a fixed temperature for the falling atoms, as the adiabatic cooling is not efficient for a non trapped cloud and the heating rate is also very low after $z_{\rm out}$. $T_h$ is about $10~\mu$K for Cs and $25~\mu$K Rb. After the remaining falling time of 36~ms (resp. 36~ms for Rb) the atomic beam has a typical standard deviation for the transverse Gaussian atomic density distribution of $\Delta r_f \approx 1~$mm for Cs and $\Delta r_f \approx 1.75~$mm for Rb, smaller than MOT2 radius, meaning that almost all the atoms are recaptured in MOT2 for both experiments. Note that this model allows to predict $\Delta r(z)$ at any position $z$, as shown on figure~\ref{fig:sizeRb}, right.

\subsection{Transfer efficiency}
\label{subsec:TransferEff}

\begin{figure}[t]
\begin{center}
\includegraphics[width=0.83\linewidth]{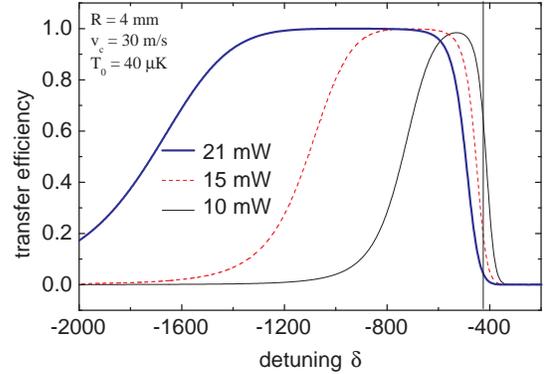}
\end{center}
\caption{(Color online) Efficiency (see text) of the pushing--guiding processes versus laser detuning $\delta$, calculated for $^{87}$Rb for the parameter of the pushing beam given in text, with different laser power 10 mW (thin solid line, black), 15 mW (dashed line, red) and 21~mW (thick solid line, blue). The maximal capture velocity in MOT2 has been fixed to $v_{\rm capture} = 30$~m/s, the initial temperature to $T_0=40~\mu$K and the MOT2 beam radius $R$ to 4~mm. The corresponding experimental values are shown on Figure~\ref{fig:detuning}, right.}
\label{fig:efficiency}
\end{figure}

\begin{figure*}
\resizebox{1\textwidth}{!}{
\rotatebox[origin=rB]{270}{
\includegraphics*[width=\linewidth]{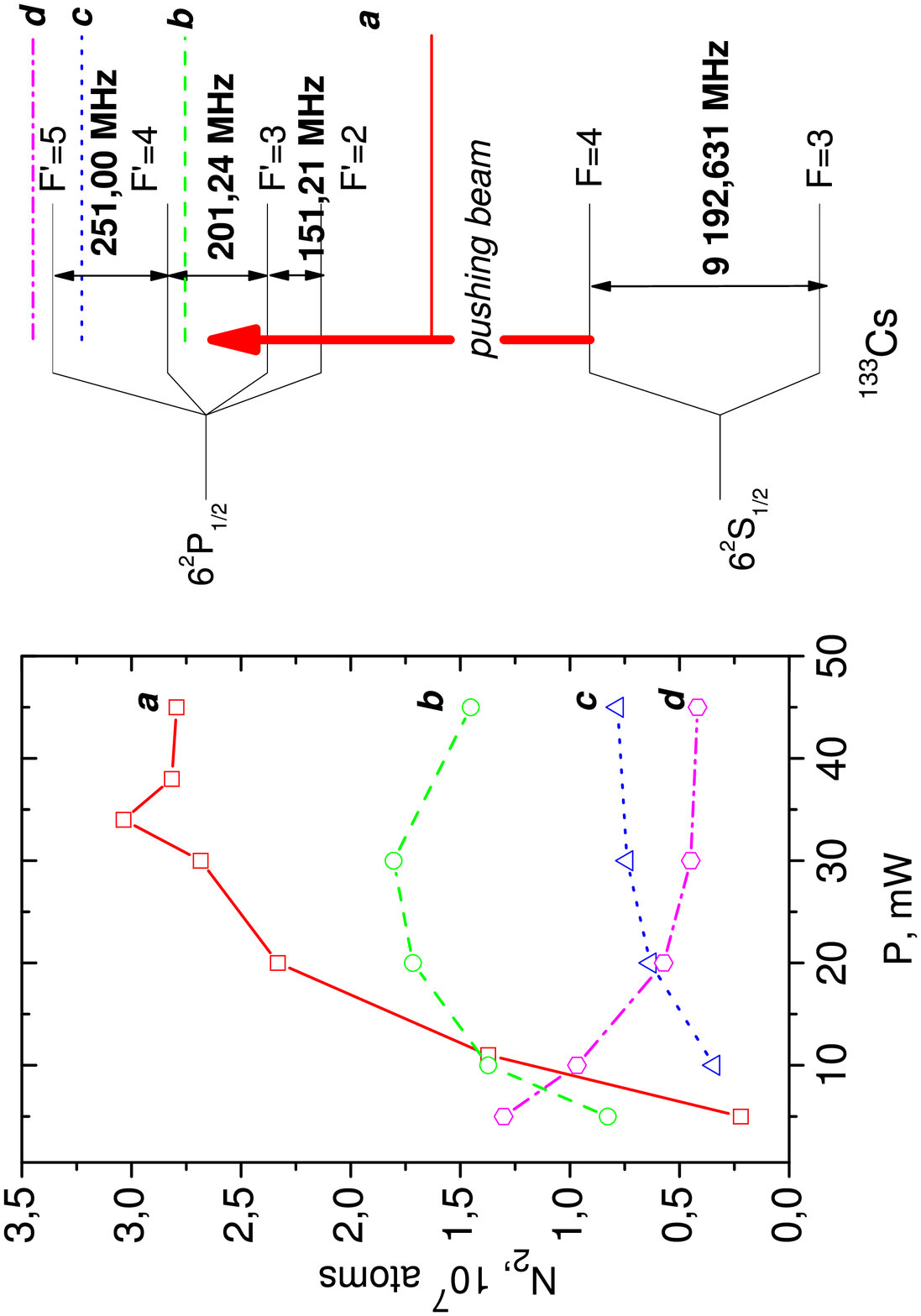}}

\rotatebox[origin=rB]{270}{
\includegraphics*[width=\linewidth]{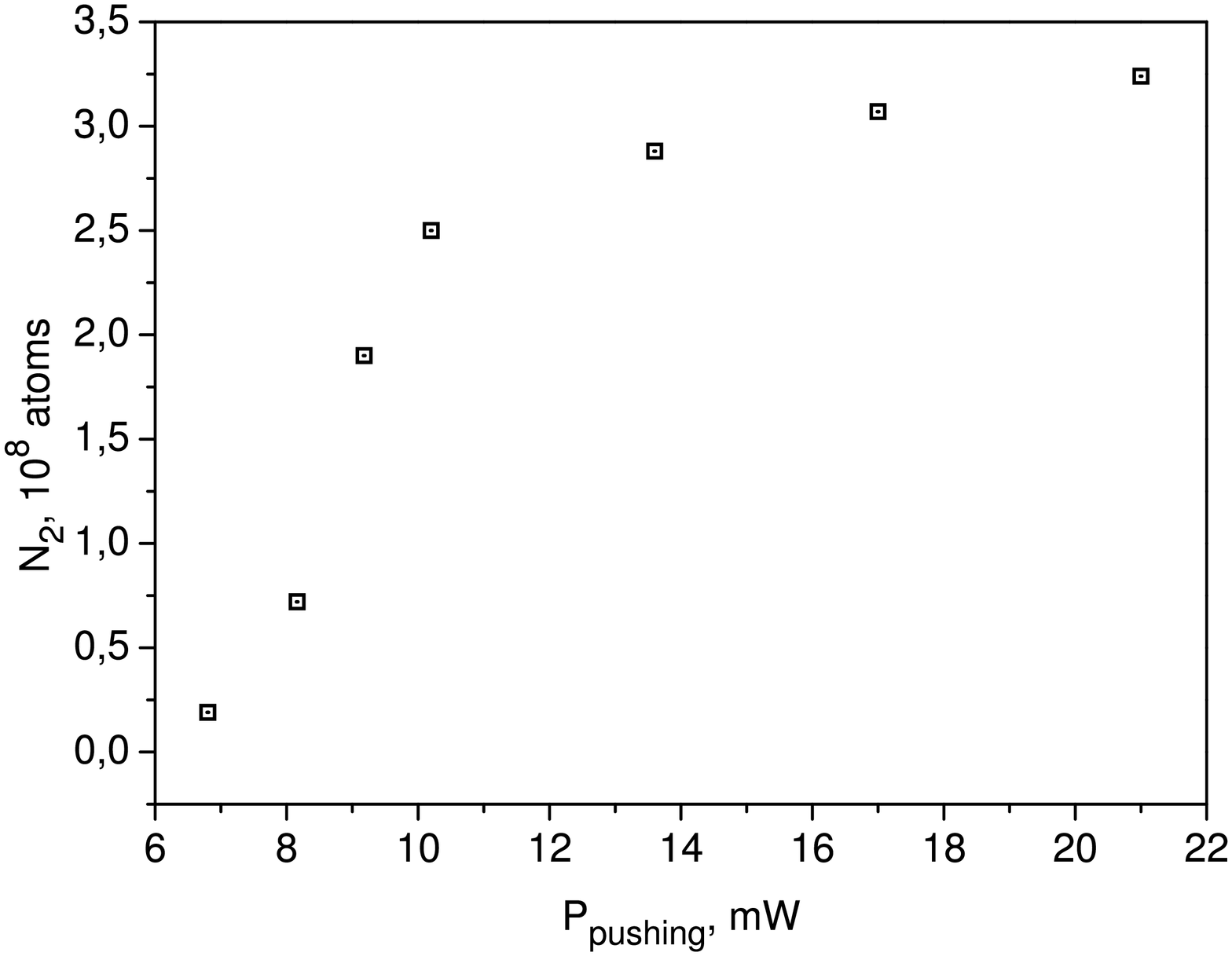}
}
}
\caption{Dependence of the number of recaptured atoms $N_{2}$ in MOT2 on the pushing beam power. Left: Cs data. The experiment is done at four different frequencies (see the diagram in the centre). Right: $^{87}$Rb data, recorded at -1~GHz detuning from the $5S_{1/2} (F=2) \rightarrow 5P_{3/2} (F^{'}=3)$  transition.}
\label{fig:power@detuning}
\end{figure*}

We come back now to an estimation of the transfer efficiency as discussed in section~\ref{subsec:2level} and presented in Figure~\ref{fig:pot_2_niv}. Within the frame of the refined model presented now, we are able to compute a transfer efficiency in the same spirit. As we have seen in the previous section, the guiding is not required until the end for the whole cloud to be recaptured. We thus retained the two following conditions: (i) the arrival velocity has to be smaller than $v_{\rm capture}$ and (ii) the cloud size must be less than the MOT2 beam waist. We then calculate the efficiency $f[\Delta r(D)/R] \times$\\ $ f[v(D)/v_{\rm capture}]$, with the function $f$ previously used in section~\ref{subsec:2level}, and plot it on Figure~\ref{fig:efficiency}. The model predicts a good efficiency in a detuning range between -0.5~GHz and -1.6~GHz, the width of the large efficiency region being reduced with a smaller laser power. These predictions have to be compared with the rubidium experimental data of Figure~\ref{fig:detuning}, right. The agreement is qualitatively good, and reproduces the main features. The two limits of the large efficiency region have different origins: On the large detuning side, the efficiency drops due to an increase in the atomic cloud size, as the guiding potential is weaker. On the lower detuning side, close to resonance, the efficiency becomes limited by the final velocity, which is larger than the capture velocity of MOT2. On this side, theory fails to predict the less measured efficiency at lower laser power, as the mean detuning $\bar{\delta}$ approach (section~\ref{sec:OptPump}) is not valid any more. In particular, the efficiency should drop to zero at the resonance with the rubidium $5S_{1/2} (F=2) \rightarrow 5P_{3/2} (F^{'}=1)$ line, situated at $\delta/2\pi = -424$~MHz and marked with a vertical line on Figure~\ref{fig:efficiency}.

\section{Experimental results}

In this section, we present the experimental study of the guiding process and compare it with the above theoretical model. The dependence of the recaptured atom number on the pushing beam parameters are first investigated. We then
measure the mean atomic velocity and the traveling time. During the experimental investigation the atom vapour pressure in MOT1 is kept constant.

\subsection{Pushing Beam Parameters}

The parameters of the pushing beam that we have experimentally optimized are its divergence, waist, power and detuning.

\paragraph{Divergence and waist}
\label{sec:DivergenceAndWaist}

In order to  optimize  the atomic beam characteristics we have first investigated the role of the laser beam waist, related to the divergence of the pushing beam and to the pushing force. It is clear that the pushing--guiding beam should diverge, to have a significant effect on MOT1 without disturbing MOT2. Moreover, this divergence provides an horizontal adiabatic cooling of the guided atoms. We have used  three different lenses ($f$= 0.75~m ; 1~m ; 2~m) to focus the pushing beam. For each lens the transfer efficiency is studied as a function of the focus distance from MOT1. The position of the lens is more critical than its focal length. The optimum is obtained with a lens $f= 2$ m for the Cs experiment (resp. $f= 1$ m for $^{87}$Rb) and distance from MOT1 $34$~cm (resp. $\approx 13$~cm), where the beam diameter on the MOT1 region is $\approx 1.3$~mm (resp. $\approx 0.6$~mm). The measured waist at the focal point is $200~\mu$m (resp. $300~\mu$m). It leads to a divergence $w_0/z_R$ of about 2~mrad (resp. 1~mrad).
 
In conclusion we found that the best transfer efficiency occurs when the pushing beam, focused before MOT1, has a diameter smaller than 1~mm in MOT1 and a divergence such  that the beam diameter at MOT2 position is less than 3 mm. In this sense our results are similar to the one found in references \cite{Wohlleben2001,Cacciapuoti2001}.

\paragraph{Power and Detuning}

\begin{figure*}
\resizebox{1\textwidth}{!}{
\rotatebox[origin=rB]{270}{
\includegraphics*[width=0.69\linewidth]{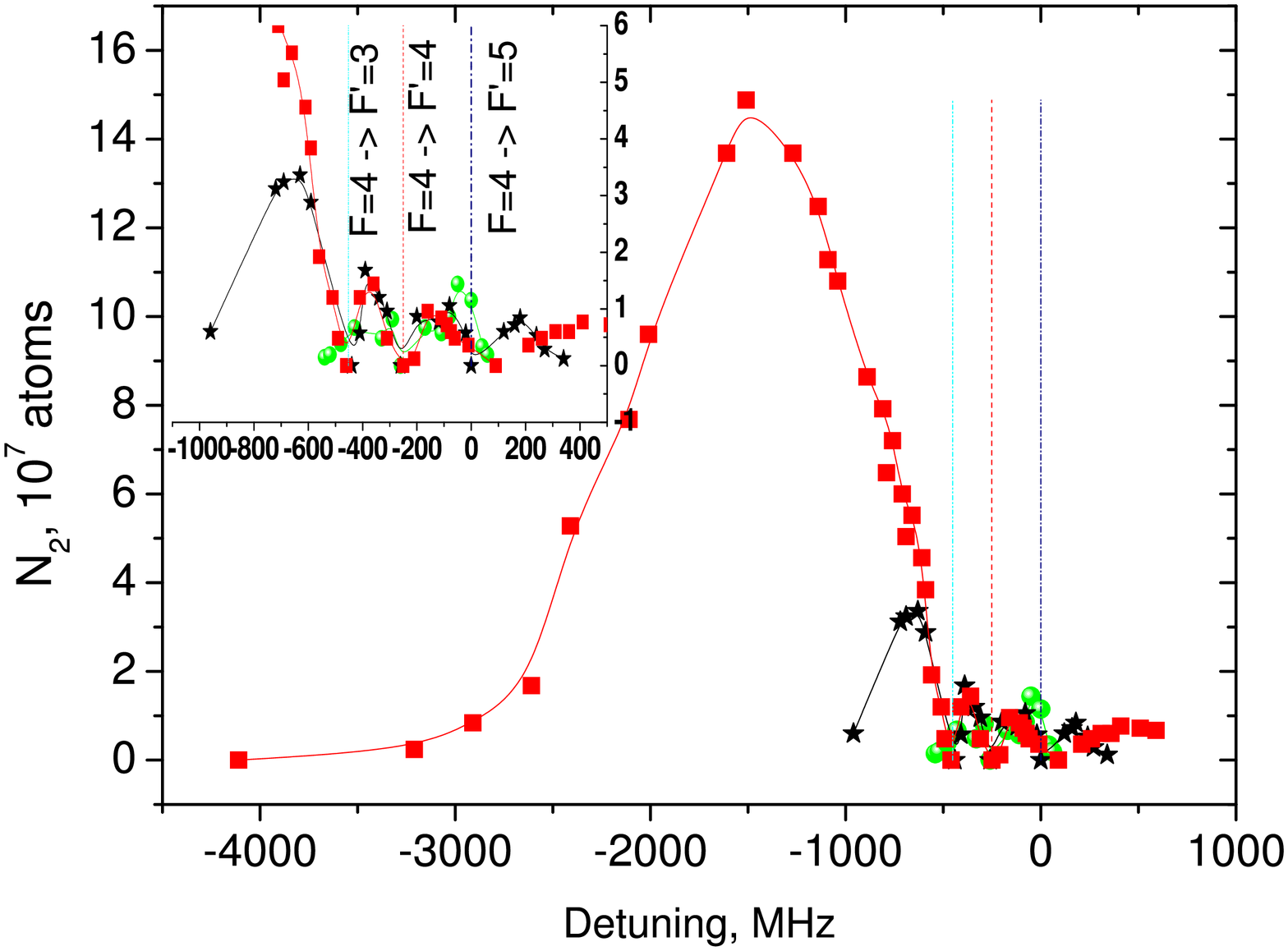}
}
\includegraphics[width=\linewidth]{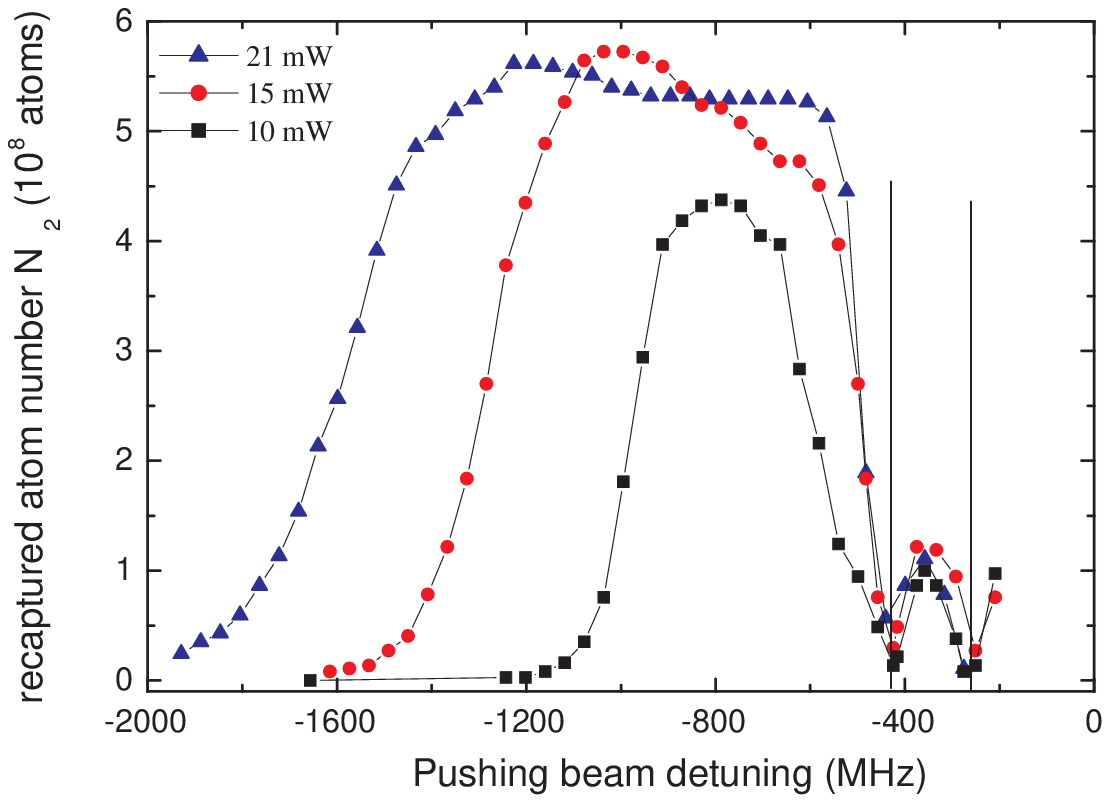}
}
\caption{(Color online) Number of atoms recaptured in MOT2 $N_2$ vs. pushing beam detuning for different optical powers. The vertical lines indicate the position of hyperfine resonance frequencies. Left: Cs data. The detuning is given with respect to the $6S_{1/2}(F=4)\rightarrow 6P_{3/2}(F'=5)$ transition. The pushing beam power is 46~mW (squares), 10~mW (stars) or 2~mW (circles). Right: $^{87}$Rb data. The frequency is measured relatively to the $5S_{1/2} (F=2) \rightarrow 5P_{3/2} (F^{'}=3)$ cycling transition. Pushing beam power: 10~mW (squares), 15~mW (circles), and 21~mW (triangles).}
\label{fig:detuning}
\end{figure*}

The recaptured number of atoms into MOT2 at different laser powers of the pushing beam and at different detunings for the two elements Cs and $^{87}$Rb is shown resp. on left and right of Figure~\ref{fig:power@detuning}. 

It is first obvious that the best experimental conditions are achieved with a laser frequency red-detuned with respect to all  atomic transitions (curve (a) in Figure~\ref{fig:power@detuning}, left). 
The transfer efficiency is larger for a red-detuned laser frequency than for the other laser frequencies due to the fact that after  leaving the MOT1 area the atoms feel the pushing light also as a guide. For such detunings, the atomic flux as well as the number of recaptured atoms $N_{2}$  in MOT2 increase when the power of the pushing light increases, and saturates at large power when all the atoms are efficiently guided to MOT2 (see also Figure~\ref{fig:power@detuning}, right). At a given detuning, an increase of the laser power leads to a  decrease of the transfer efficiency, due to an excessive final velocity, a strong perturbation of both MOTs, and a large heating of the atoms.

In order to optimize the  conditions for the atomic beam, the influence of the detuning of the pushing light was investigated  in more details (see Figure~\ref{fig:detuning} left and right resp. for Cs and $^{87}$Rb). For a frequency close to  resonance (corresponding  to the best conditions found in reference~\cite{Wohlleben2001,Cacciapuoti2001}) the number of  recaptured atoms into MOT2 is much smaller than the one we could achieve with a much more red-detuned light and a higher power. In conclusion we find that the best loading of MOT2 is at highest possible power of the pushing laser beam and, given this power, at the value of red detuning optimizing the flux.

\subsection{Atomic beam velocity}
\label{sec:BeamVelocity}
\begin{figure}
\includegraphics*[width=0.9\linewidth]{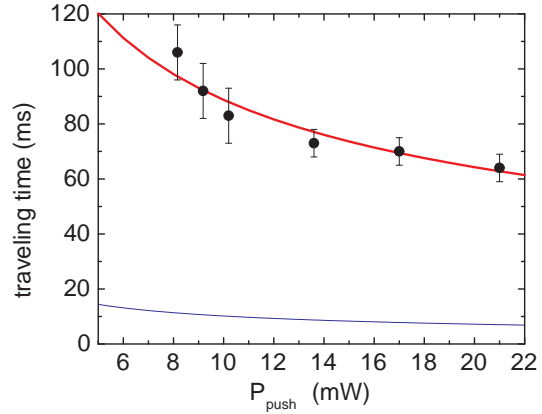}
\caption{(Color online) Experimental (points) and theoretical results (solid lines) for the traveling time $\Delta t$ between MOT1 and MOT2 for different pushing beam powers. The beam is red-detuned by 1 GHz from the cycling transition of $^{87}$Rb, $5S_{1/2} (F=2) \rightarrow 5P_{3/2} (F^{'}=3)$. The theoretical calculations are done for both the two-level model approximation (blue lower curve) and for the more detailed model described above (red upper curve). In the calculations the radius of MOT1 trapping region is 10~mm. }
\label{fig:timedelay}
\end{figure}

For a high recapture efficiency, a relatively slow and collimated atomic beam is required (see section~\ref{subsec:Recaptured}). After the pushing and guiding process, the atoms reach MOT2 within a time delay $\Delta t$. This time has been  measured in two different ways. First, one can record the MOT2 fluorescence after having suddenly removed the atoms in MOT1 (the MOT1 laser beams are stopped by a mechanical shutter). In this case, one observes the delay after which the number of atoms in MOT2 starts to drop. The second method consists in pulsing the pushing beam through a permanently loaded  MOT1. Both methods lead to the same result $\Delta t \approx 130$ ms for the Cs experiment at 63~mW power. In the Rb experiment presented in Figure~\ref{fig:timedelay}, the measured time delay  as a function of the pushing beam power is obtained by using the second method. A similar dependence on the pushing beam power is observed in the cesium experiment. The two--level model is not sufficient to describe accurately the atomic beam velocity, the predicted transfer time $\Delta t$ being by far too short (see Figure~\ref{fig:timedelay}, lower curve). On the contrary, the theoretical model presented in section \ref{sec: Pushing force}, Equation~(\ref{eqn:vel}) describes well the experimental results as demonstrated on Figure~\ref{fig:timedelay}. From the model, we also deduce the final longitudinal velocity of the atomic beam $v \approx 5.5$~m/s (resp. 12.6~m/s for Rb). Note that this final velocity is not very different from the mean velocity $D/\Delta t$, as the acceleration stage takes place essentially in the MOT1 zone, where the atoms remain in the $F+1$ state thanks to the repumping MOT beams.

\section{Conclusion}

In our work we have studied a very efficient setup to transfer cold atoms from a first MOT  to a second one. Our setups have a  similar geometry to the ones described in references~\cite{Wohlleben2001,Cacciapuoti2001}, but due to the higher laser power (tens of mW) we could achieve a partial dipolar guide for the atoms at a larger detuning (1~GHz typically). As a result, the  mean longitudinal velocity of the atomic beam is lower (4.3-12 m/s) than in these previous experiments (15~m/s). Moreover, thanks to the lower sensibility of the method to the frequency of the pushing laser, its frequency does not need to be locked (see for instance the detuning dependence in Figure~\ref{fig:detuning}, right) and the setup is much more robust to small mis-alignments of the pushing beam. The atomic flux is limited only by the number of atoms loaded into MOT1. We estimated the transfer efficiency to MOT2 which is about $70~\%$ for the $^{133}$Cs experiment and about $50~\%$ for the $^{87}$Rb experiment.

We used a two-level system model to describe the pro\-cesses during the atomic transfer. A good qualitative agreement between theory and experiment was found. The transfer efficiency is maximum for a large red detuning, and this maximum efficiency increases with the laser power. A more detailed discussion of the pushing, guiding and recapture processes is presented for a better understanding of the atomic transfer between the two traps. Our theoretical description, which takes into account the optical pumping, the pushing force and the guiding potential nicely reproduces the experimentally observed traveling time.

In conclusion, we experimentally described and theoretically modelled a method to transfer cold atoms between two traps. 
Two different setups lead qualitatively to the same optimized parameters -- a large laser power (tens of mW), $\approx$ 1~GHz detuning, $300~\mu$m waist. The implementation of this technique in our setups brought in both cases a much better stability and improved loading efficiency, with respect to the use of a near resonant laser beam.

\begin{acknowledgement}
This work was partially supported by European RTN networks COMOL, QUACS, FASTNet and Atom Chips. LAC and LPL are members of IFRAF. LAC is UPR~3321 of CNRS associ\'ee \`a l'Universit\'e Paris-Sud and member of LUMAT FR~2764, and LPL is UMR~7538 of CNRS and Universit\'e Paris 13. We are very grateful to Brigitte Mercier for giving us the seminal idea of this pushing--guiding scheme. The authors also thank Laurent Vernac for helpful discussions on optical pumping and for providing us a computing code on the calculation of the populations in the magnetic sublevels.
\end{acknowledgement}


\begin{thebibliography}{40}
\bibitem{note1} This population ratio may be adjusted by using an elliptically polarized laser beam. It may be a way to optimize experimentally the final atomic beam velocity. In the model, we restrict ourselves to a linear polarization for sake of simplicity.
\bibitem{note2} In this model, we have neglected the heating term related to the fluctuations of the guiding force when the atom jumps between the two hyperfine states. However, this contribution is smaller by about a factor $\left[\lambda \delta/(2 \pi w \Gamma)\right]^2$ than the spontaneous scattering term, that is more than two orders of magnitude everywhere along the guide.
\end{thebibliography}
\end{document}